# ENERGY SPECTRUMS OF SLOWED NEUTRONS IN HETEROGENEOUS URANIUM-CARBON AND THORIUM-CARBON FISSION MEDIA

V.V. Lavrukhin, S.A. Chernezhenko, M.R. Shcherbyna, V.O. Tarasov*, E.Yu. Zhelezko

*Department of Theoretical and Experimental Nuclear Physics, Odessa National Polytechnic University, Shevchenko av. 1, Odessa 65044, Ukraine*

**Abstract**

Using the Geant4 and OpenMC Monte Carlo codes, the energy spectra of moderated neutrons were studied in homogeneous uranium dicarbide and uranium dioxide media, as well as in heterogeneous uranium-carbon and thorium-carbon channel-type fission structures with various compositions and lattice parameters. In homogeneous uranium dicarbide the spectral maximum lies in the 20–50 keV range, which is important for the design of a prototype fast single-channel reactor operating in the traveling-wave fission mode with a soft fast-neutron spectrum. A heterogeneous thorium-carbon medium that forms a thermal neutron spectrum has been identified, enabling the realization of a traveling-wave mode of neutron-nuclear fission on thermal neutrons. By varying channel lattice parameters and introducing burnable absorbers (Cd, In), the possibility of forming a superthermal neutron spectrum (1–20 eV) in a heterogeneous uranium-carbon medium was demonstrated. The results are useful for the development of nuclear transmutation reactors and next-generation reactors operating in the traveling-wave mode.

## I. Introduction

An important direction in the development of nuclear energy, based on nuclear chain reactions involving the fission of heavy nuclides by neutrons, is the development of nuclear transmutation reactors, in which the formation of minor nuclides is minimized, thereby ensuring the biocompatibility of nuclear energy with the human environment. Research is being conducted into the possibilities of minimizing the formation of minor nuclides by moderating the energy spectra of fast nuclear reactors through the placement of specialized carbon assemblies in the core, for example, [1]. In addition, new-generation nuclear reactors are being developed that operate in a traveling-wave mode of neutron-nuclear fission not only on fast neutrons but also on intermediate and thermal neutrons (e.g., [2–10]), which is also consistent with the strategy for creating transmutation reactors. It is important to note that the traveling-wave mode of neutron-nuclear fission in uranium fission media can be achieved only with fast and intermediate neutrons (e.g., [2–6]), whereas in thorium fission media, it can also be achieved with thermal neutrons. It should also be noted that nuclear reactors operating in the traveling-wave mode of neutron-nuclear fission are so-called “inherently safe” reactors; that is, they are based on new physical principles (self-regulation of the chain fission reaction) that address the main challenges of ensuring the safe operation of nuclear reactors.

In this work, using the Geant4 [11] and OpenMC [12] software codes, which implement the Monte Carlo method, a study was conducted of the energy spectra of moderated neutrons in various homogeneous and heterogeneous uranium and thorium fission structures with different compositions and channel lattice parameters. For example, neutron spectra were investigated in homogeneous uranium fissile media (uranium dicarbide and uranium dioxide), as well as in heterogeneous uranium

* Corresponding author e-mail: vtarasov@ukr.net

carbon (neutron moderator—carbon) and thorium-carbon (moderator—carbon) channel-type fission media.

The results obtained may be useful in the development of nuclear transmutation reactors—that is, to minimize the production of minor nuclides by the nuclear power industry, as well as to determine the composition and structure of the breeding medium in next-generation nuclear reactors operating in the traveling-wave mode of neutron-nuclear fission.

A heterogeneous channel-type thorium-carbon breeding medium has been characterized, in which a thermal neutron spectrum is formed; consequently, it is possible to realize a traveling-wave mode of neutron-nuclear fission on thermal neutrons in such a thorium-carbon breeding medium [6–10].

## II. Neutron moderation spectra in homogeneous fission media consisting of uranium dicarbide and uranium dioxide

Today, the development of nuclear reactors operating in the traveling-wave mode of nuclear fission on fast neutrons faces the challenge of ensuring the radiation resistance of the core structural materials at a level of 500 ZNA (displacements per atom) [2]. Several ideas have been proposed to address this problem [3, 4, 13]. Paper [13] presents a conceptual design of a prototype fast single-channel reactor operating in a traveling-wave fission mode with a soft fast-neutron spectrum. The problem of radiation resistance of structural materials in this experimental reactor design is addressed based on proposals previously published by the authors [3, 4]. Specifically, a traveling wave of nuclear fissions is implemented using neutrons with a moderated fast spectrum (the spectral peak lies in the 20–50 keV range), which reduces the neutron flux on structural materials by more than an order of magnitude. It is proposed to use uranium dicarbide in the form of a cylinder—which is homogeneous with respect to the neutron field—as the nuclear fuel through which the fission wave propagates, but this proposal, which was based on studies of neutron moderation spectra in homogeneous uranium-carbon mixtures [14–16], required further investigation of the spectra of moderated neutrons in a homogeneous fission medium consisting of uranium dicarbide.

Recall that neutrons in neutron-multiplying (fission) media are produced in a chain reaction as a result of nuclear fission reactions of fissionable nuclides (uranium-235, plutonium-239, and uranium-233) with energies described by the fission spectra of the corresponding fissile nuclides. Moreover, the neutron fission spectra of these fissile nuclides are similar to one another and have comparable characteristics: the main range of neutron energies is from 0 eV to 10 MeV; the most probable neutron energy is 1 MeV, and the average energy is 2 MeV [17, 18]. As a result of the interaction of neutrons with the multiplying medium, the neutron energy spectrum changes and is described by the neutron moderation spectrum. The neutron moderation spectrum depends on the composition and structure of the neutron multiplying medium, as well as on its temperature.

Therefore, using the OpenMC software code [12] and the Monte Carlo method, calculated neutron moderation spectra were obtained (the probability density function $\varphi(E)$, where $\varphi(E)dE$ is the probability density function that a neutron has kinetic energy in the energy range from E to E+dE, as a function of E), as well as the neutron flux densities, which are emitted by an isotropic, uniformly distributed source with a fission spectrum in a medium consisting of uranium dicarbide (15% enriched in uranium-235) and uranium dioxide (15% enriched in uranium-235). Calculations of the neutron spectra were performed for a splitting medium in the form of a cylinder with a diameter of 20 cm and a height of 100 cm, located in a vacuum. Neutron moderation spectra and neutron flux densities were obtained for four points located in the cross-sectional plane of the cylinder at a height of 50 cm, at various distances from the cylinder's axis: point A is located on the cylinder's axis, point B is located 10 cm from the cylinder's axis, point C is located on the cylinder's surface, and point D is located in a vacuum at a distance less than the distance to the extrapolated boundary. The calculation results are shown in Figures 1 and 2, respectively, for the energy spectra of moderated neutrons and for the neutron flux density at points A, B, C, and D in uranium dicarbide. Similarly, Figures 3 and 4 show the calculated neutron moderation spectra and neutron flux densities, respectively, in uranium dioxide.

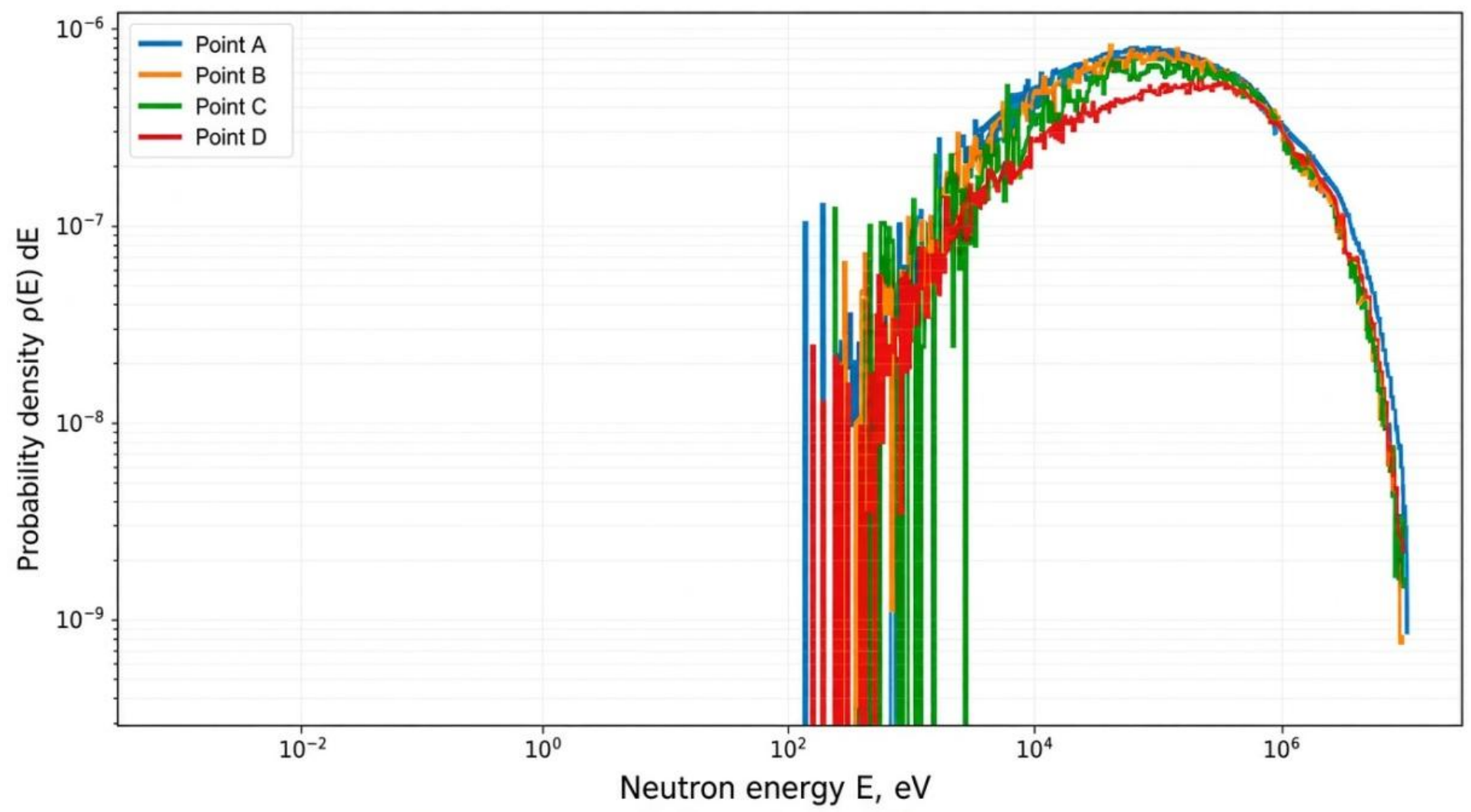

*Fig. 1. Calculated neutron moderation spectra in uranium dicarbide for four points located in the neutron cross-section plane at a height of 50 cm, at various distances from the cylinder axis: point A is located on the cylinder axis, point B is located 10 cm from the cylinder axis, point C is located on the cylinder surface, and point D is located in a vacuum at a distance less than the distance to the extrapolated boundary.*

It is worth noting some peculiarities in the presentation of the calculation results shown in Figures 2 and 4 regarding neutron flux densities, which were obtained using the OpenMC software code [12]. The neutron flux densities were obtained and are presented in Figures 2 and 4 as relative values, since they are normalized to the number of neutrons used by the software code for the given Monte Carlo calculation (for the calculations shown in Figures 2 and 4, each calculation tracked 200, million neutron histories, of which 160 million neutrons were used for statistical estimates of the neutron spectra and flux densities), and are normalized to the neutron energy discretization interval, since a non-uniform neutron energy discretization scale is used in the calculations. Essentially, this is the energy-dependent neutron flux density spectrum in the given medium.

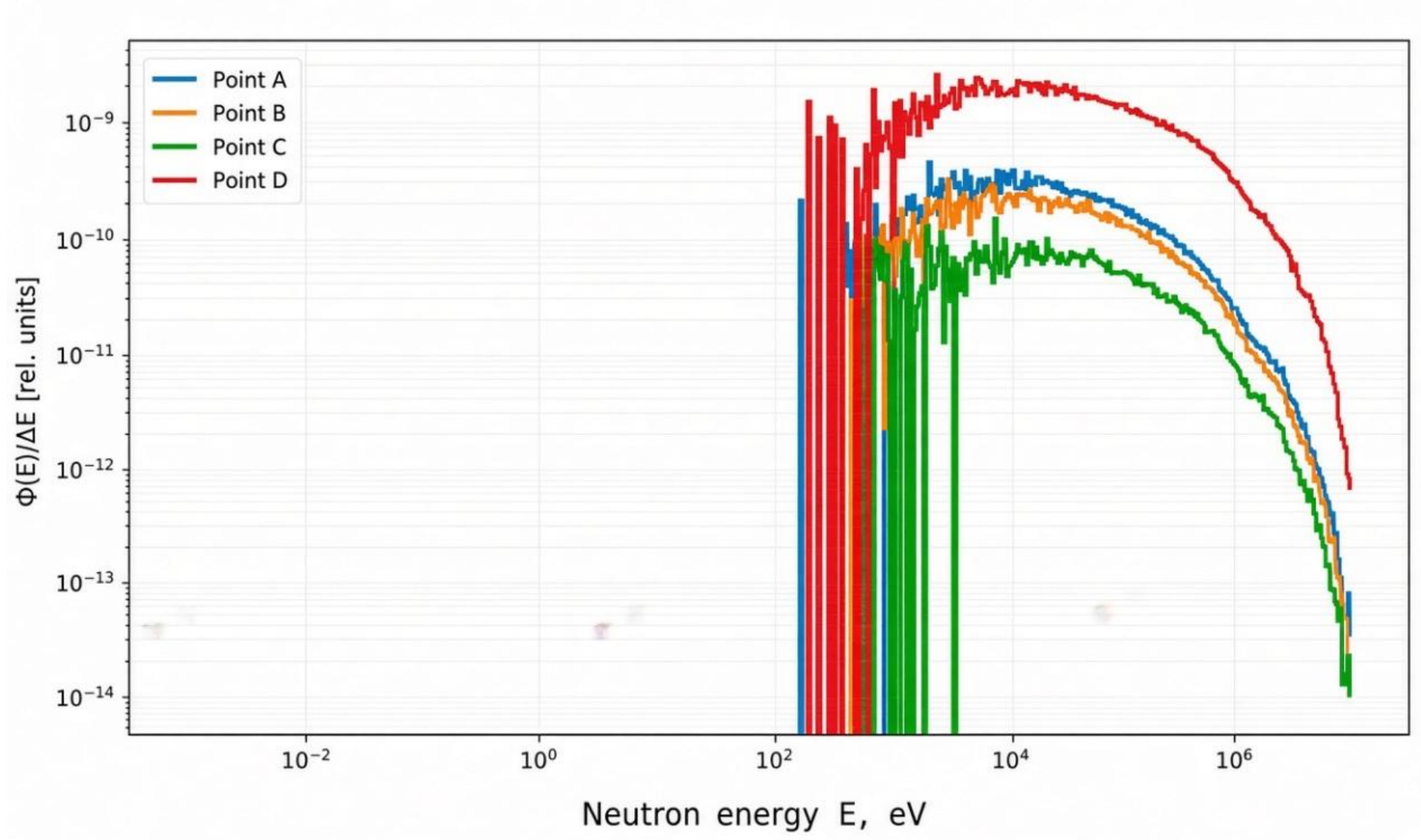

*Fig. 2. Calculated neutron flux densities in uranium dicarbide for four points located in the neutron cross-section plane at a height of 50 cm, at various distances from the cylinder axis: point A is located on the cylinder axis, point B is 10 cm from the cylinder axis, point C is located on the cylinder surface, and point D is located in a vacuum at a distance less than the distance to the extrapolated boundary.*

Thus, the additional studies conducted and their results, presented in Figures 1 and 2, refine and confirm the results of [13–16], and also show that the maximum of the neutron moderation spectra lies in the range of 20–50 keV. This is particularly evident in Figure 2 for the neutron flux density.

For a comparative analysis, Figures 3 and 4 show the calculated neutron moderation spectra and neutron flux density, respectively, in uranium dioxide. A comparative analysis of these results shows that the rapid spectrum in uranium dioxide does not undergo the same softening as in dicarbide.

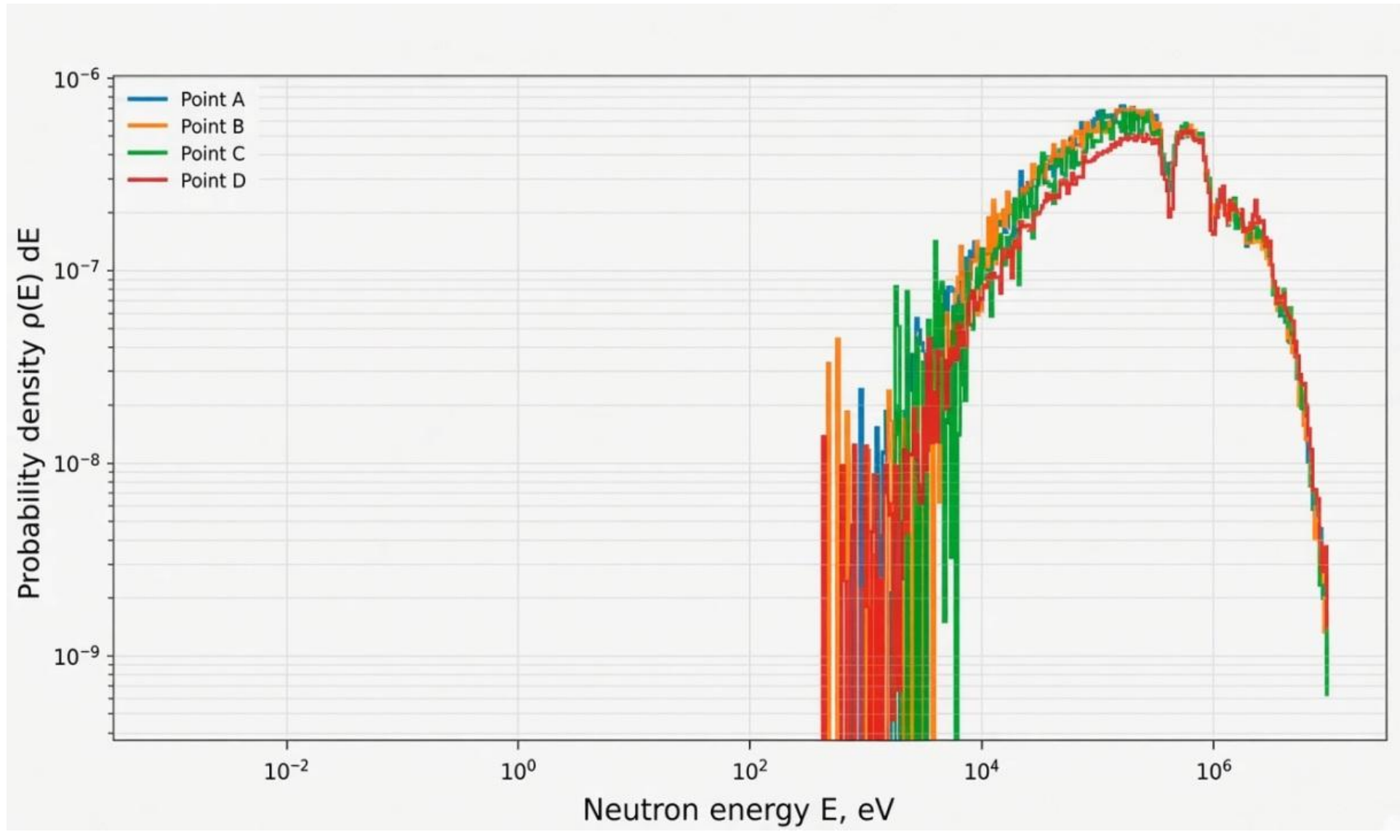

*Fig. 3. Calculated neutron moderation spectra in uranium dioxide for four points located in the neutron cross-section plane at a height of 50 cm, at various distances from the cylinder axis: point A is located on the cylinder axis, point B is located 10 cm from the cylinder axis, point C is located on the cylinder surface, and point D is located in a vacuum at a distance less than the distance to the extrapolated boundary.*

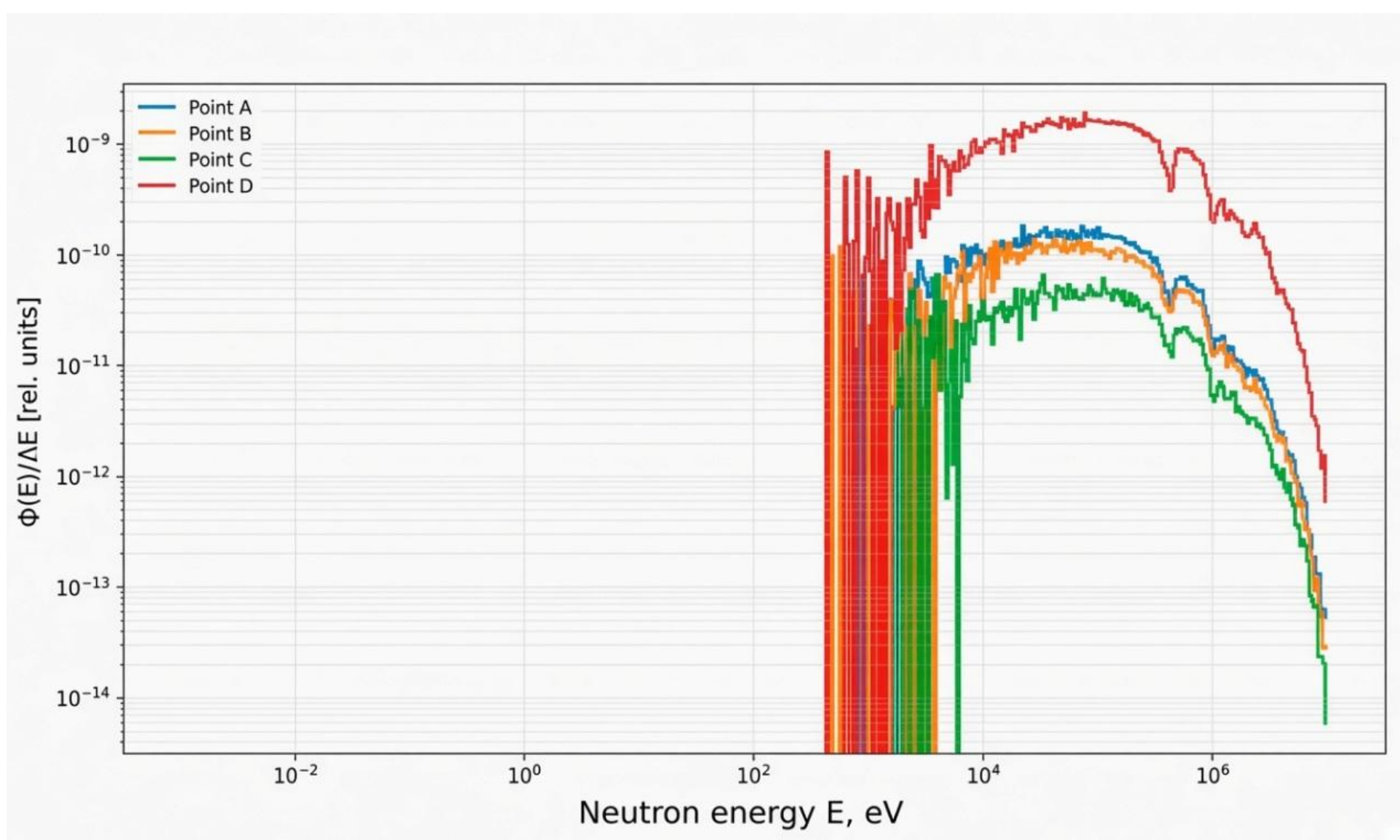

*Fig. 4. Calculated neutron flux densities in uranium dioxide for four points located in the neutron cross-section plane at a height of 50 cm, at various distances from the cylinder axis: Point A is located on the cylinder axis, point B is located 10 cm from the cylinder axis, point C is located on the cylinder surface, and point D is located in a vacuum at a distance less than the distance to the extrapolated boundary.*

## III. Neutron moderation spectra in heterogeneous uranium-carbon moderating media with different compositions and channel lattice parameters

In this work, using the Geant4 [11] and OpenMC [12] software codes, which implement the Monte Carlo method, studies were conducted on the energy spectra of slowing neutrons in various heterogeneous uranium-based media with different compositions and channel lattice parameters.

To calculate the neutron moderation spectra, a heterogeneous uranium partitioning structure was selected, analogous to the structure of the heterogeneous core of an RBMK channel-type nuclear reactor [18], as shown in Fig. 5. This choice of a heterogeneous uranium fission structure is primarily driven by the need to determine the structure and composition of the heterogeneous core for a nuclear reactor operating in the traveling-wave mode of neutron-nuclear fission on superthermal neutrons (the neutron spectrum peak lies in the range of 1–7 eV) in uranium reactor fuel [3], and the geometry of the core fuel elements in the form of long cylinders is suitable for implementing the traveling-wave fission mode. Second, it is due to the fact that a structure analogous to that of the heterogeneous core of an RBMK channel-type nuclear reactor [18] can be treated during calculations as a test heterogeneous uranium fission structure to verify the correctness of the modeling program's computational scheme, since it is known that the spectrum of slowed neutrons in the RBMK is thermal (following a Gaussian distribution with a mean energy—the spectrum maximum—corresponding to thermal energies of 0.025 eV, and a standard deviation of ≈ ± 0.5 eV).

The computational heterogeneous partitioned structure shown in Fig. 5 and Fig. 6 represents a simple grid of fuel channels with a diameter of 4 cm and a grid spacing of 30 cm (simplified within acceptable limits in the authors' opinion, but as close as possible to the fuel channel lattice in an RBMK reactor, which has a complex lattice [18]), arranged within a pyrocarbon block, which serves as the primary moderator for neutrons emitted by the fuel channels. The fuel channels consist of two tubes made of structural reactor metals that weakly absorb neutrons (e.g., zirconium or aluminum alloys) and are arranged coaxially; their effect on the neutron spectrum was neglected. The inner tube, with a diameter of 2 cm, is filled with fissile uranium material (uranium-238 enriched to 5% uranium-235). The volume between the inner and outer tubes of the fuel channel was assumed to be filled with water during the calculation, corresponding to the heat removal system of an RBMK reactor.

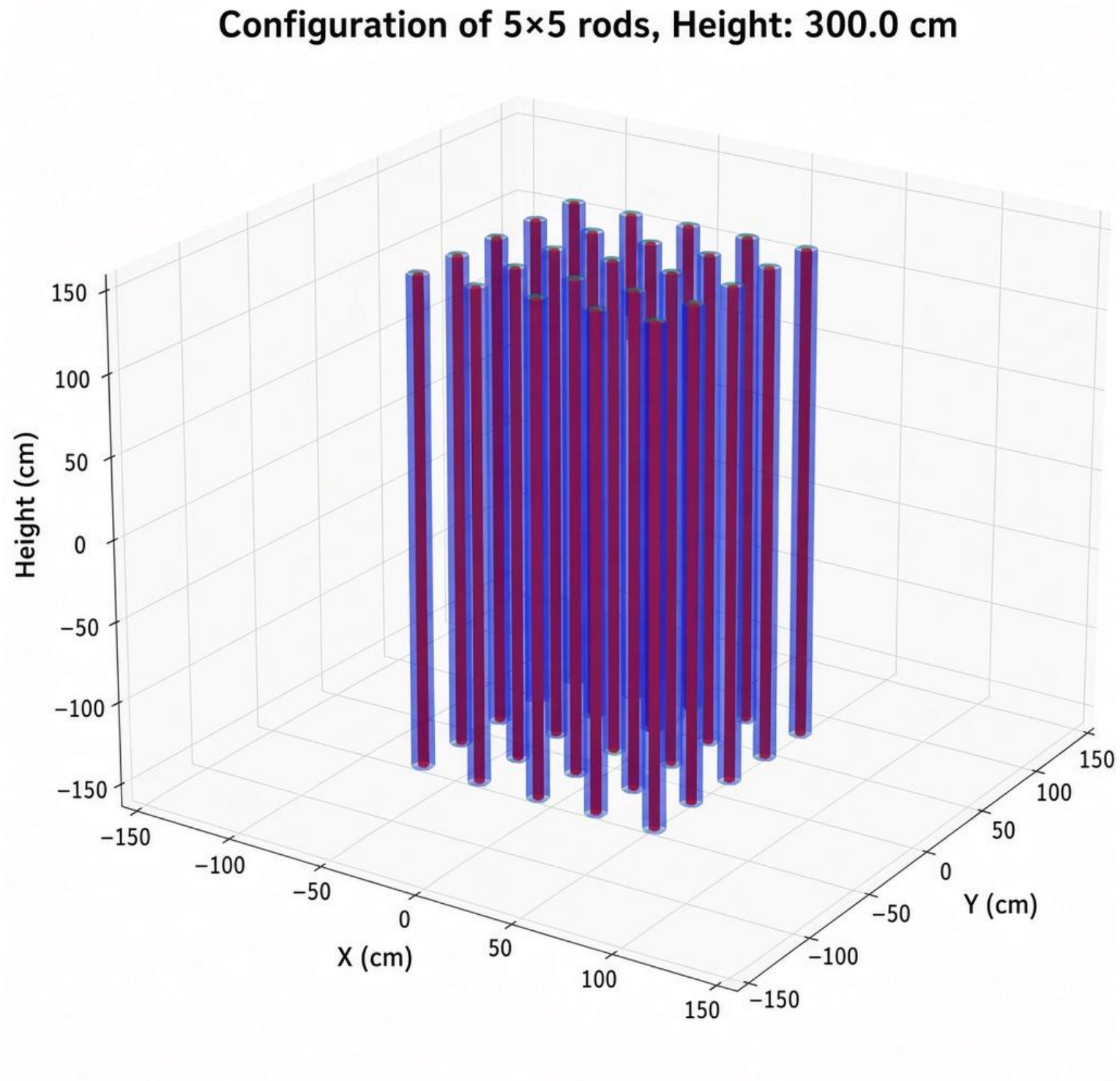

*Fig. 5. 3D structure of fuel channels in a heterogeneous neutron-multiplying medium.*

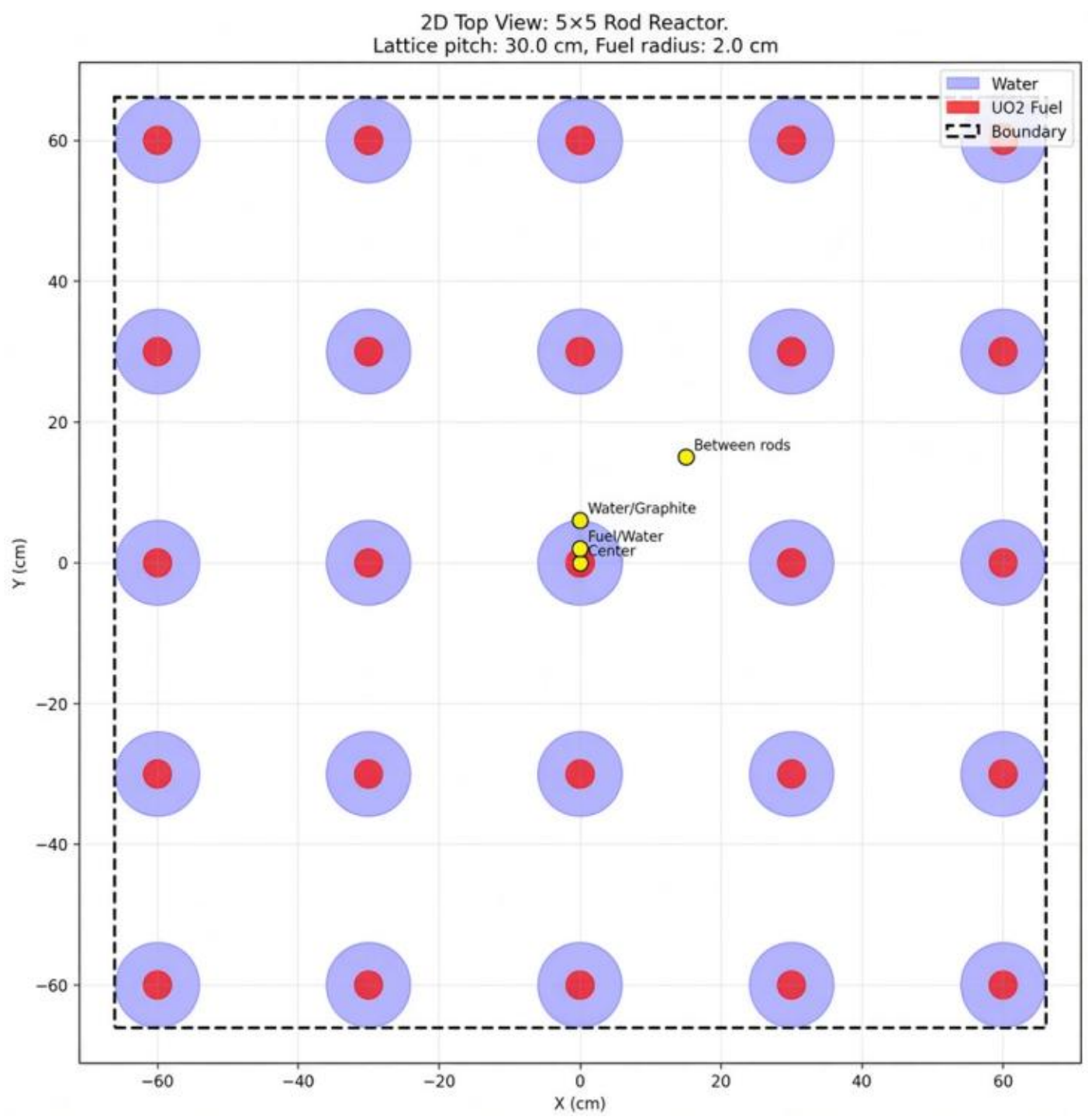

*Fig. 6. Cross-section of the 3D structure of the fuel channels of the heterogeneous neutron-multiplying medium shown in Fig. 5, and the points for which the neutron moderation spectra are shown in Fig. 7.*

Fig. 7 shows the calculated neutron moderation spectra for several points in the heterogeneous uranium fuel assembly structure depicted in Figs. 5 and 6. Note that for this calculation, the thickness of the water layer in the fuel channel was set to 1 cm, and the wall thickness of the two fuel tubes was set to 0.5 cm. Point 1 (Fig. 6) is a point located in the midplane of the heterogeneous structure and perpendicular to the fuel channels at the center of the central fuel rod; Point 2 (Fig. 6) is a point located at the interface between the fuel channel and the carbon moderator of the central rod; Point 3 (Fig. 6) is a point located in the same plane as the two previous points and within the carbon medium between the two rods.

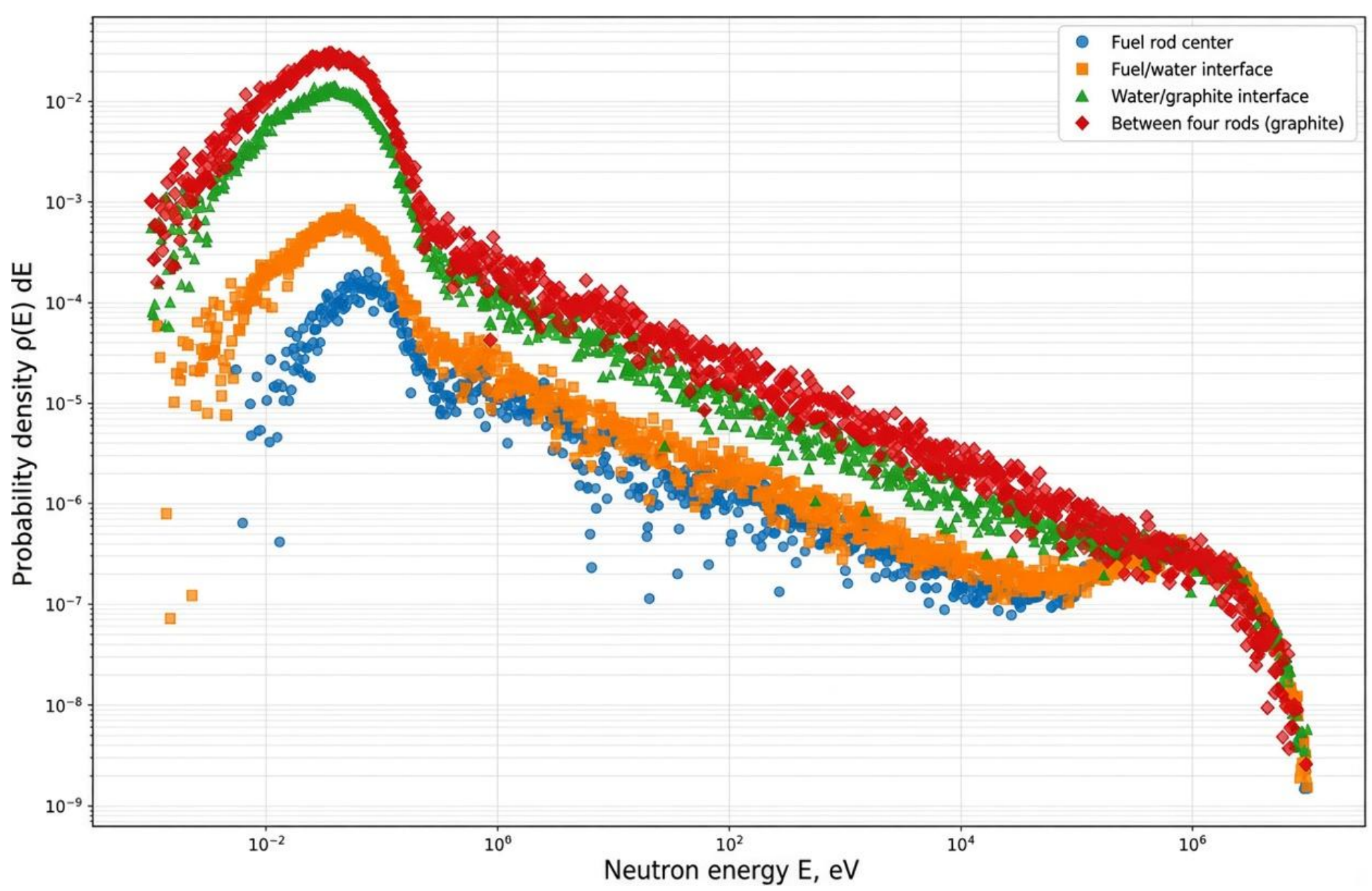

*Fig. 7. Calculated neutron energy spectra for a lattice with a pitch of 30 cm and a fuel channel diameter of 4 cm (an inner tube with a diameter of 2 cm filled with uranium-238 enriched to 5% in uranium-235).*

The calculated neutron moderation spectra shown in Fig. 7 indicate that the neutron spectra are thermal—that is, identical to those in an RBMK reactor—and it can be concluded that the validation of the modeling program's calculation scheme for neutron moderation spectra was successfully completed.

Since the main objective of the study was to determine the structure and composition of a heterogeneous core for a nuclear reactor operating in a traveling-wave mode of neutron-nuclear fission on superthermal neutrons (1–7 eV) in uranium reactor fuel, further calculated neutron moderation spectra were obtained for various parameters of the fuel channel lattice and their diameters, as well as for different compositions of fissile fuel.

Since, due to the limited scope of this article, we cannot present all the results of these studies, we will note only that for lattice parameters of 15 cm, 10 cm, and 8 cm, the spectrum changes, but the spectrum maximum is located in the thermal region. With a further decrease in the lattice parameter, the neutron spectrum increasingly approaches the fast spectrum.

Figs. 8 and 9 show, as examples, calculated moderation spectra for a heterogeneous neutron-multiplying medium similar to that shown in Figs. 5 and 6, but with different lattice and fuel channel parameters. In Fig. 8, the lattice parameter is 12 cm and the diameter of the uranium dioxide fuel channel is 1 cm, while in Fig. 9, the lattice parameter is 6 cm and the diameter of the uranium dioxide fuel channel is 1 cm.

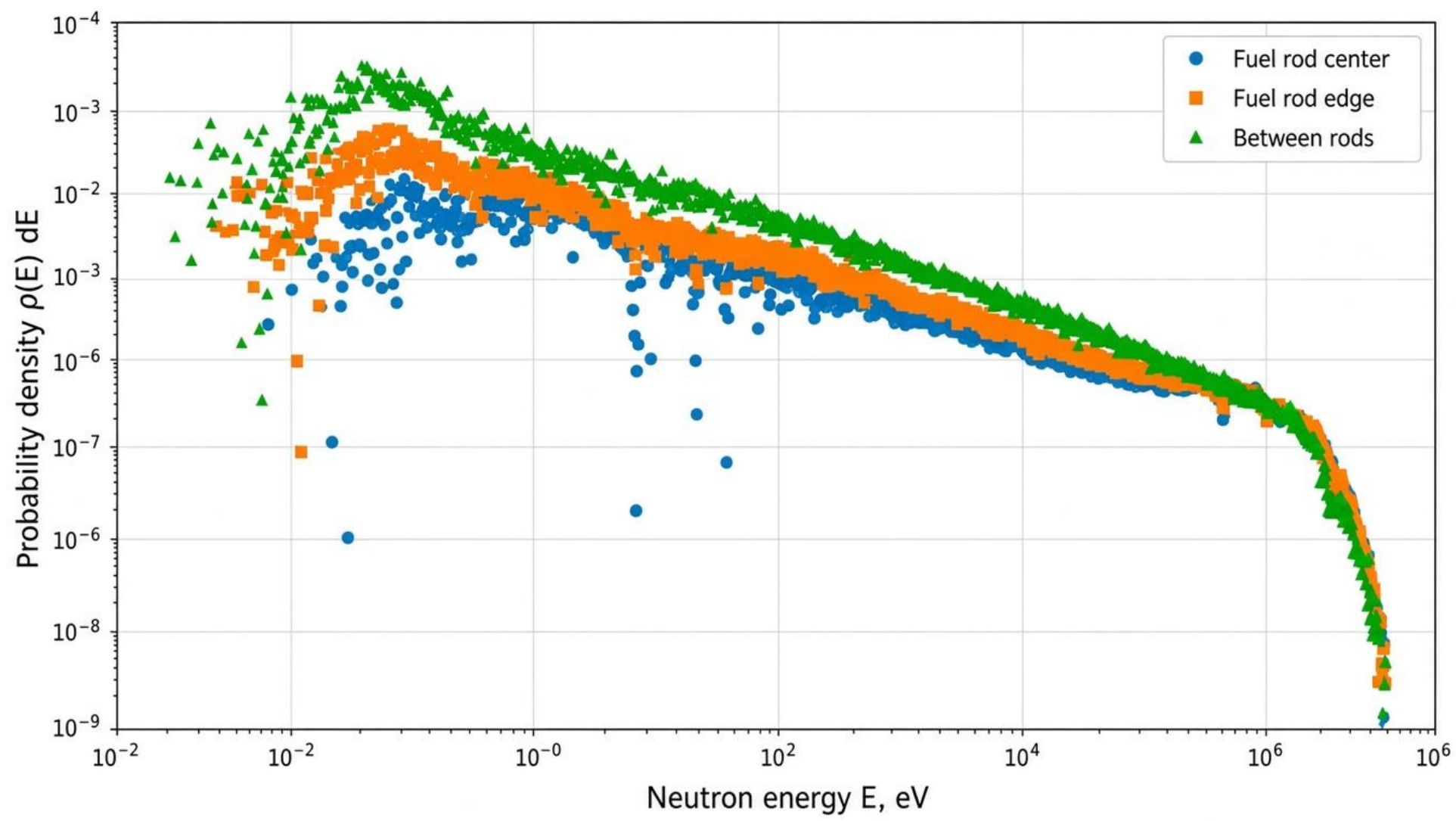

*Fig. 8. Calculated neutron energy spectra for a lattice parameter of 12 cm and a fuel channel diameter of 1 cm (an inner tube with a diameter of 1 cm filled with uranium dioxide enriched to 5% uranium-235).*

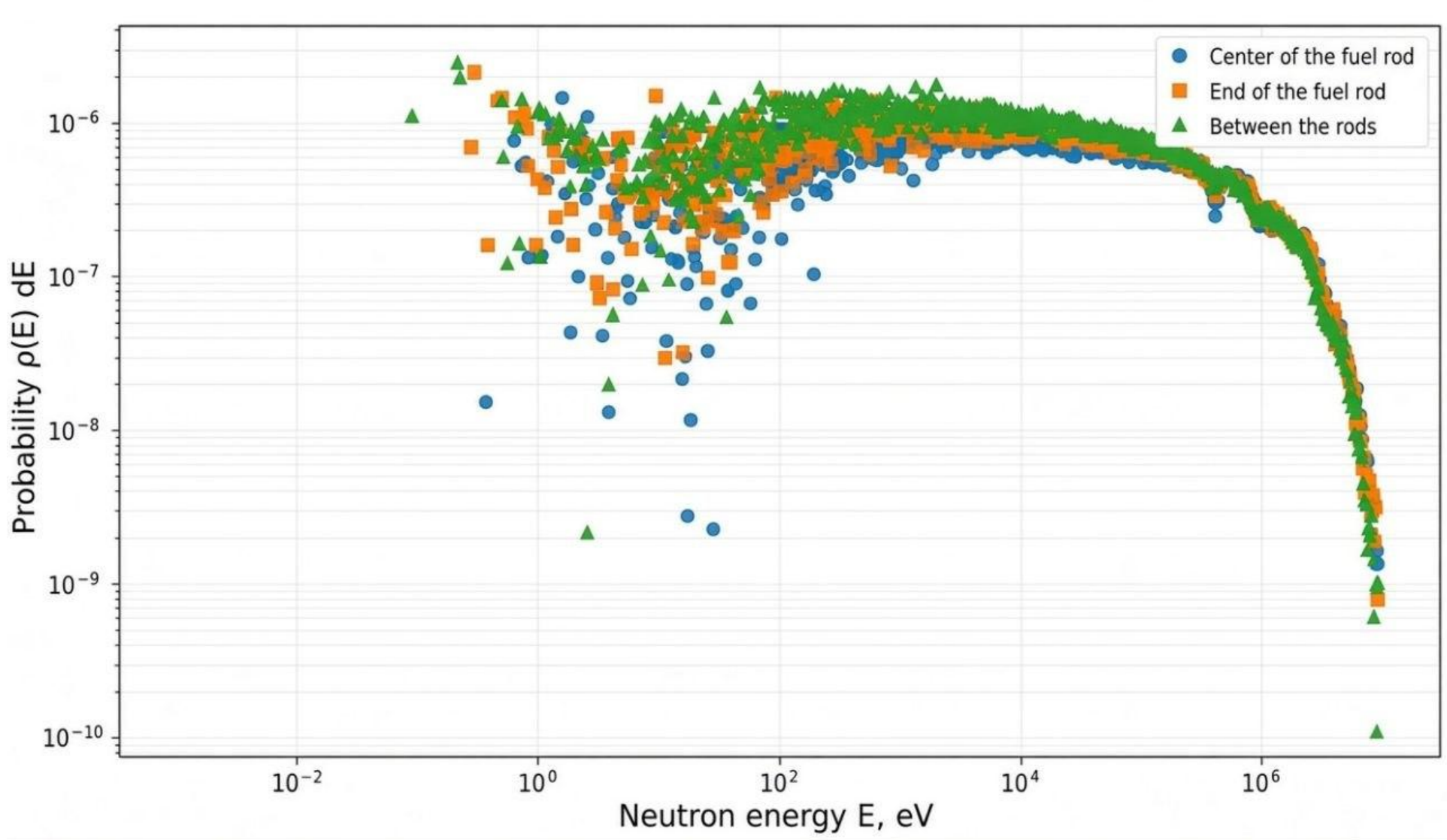

*Fig. 9. Calculated neutron energy spectra for a lattice parameter of 6 cm and a fuel channel diameter of 1 cm (an inner tube with a diameter of 1 cm filled with uranium dioxide enriched to 5% in uranium-235).*

The results presented in Figures 8 and 9 clearly demonstrate the transformation of the neutron spectrum from a thermal to a fast spectrum as the moderator layer in a heterogeneous neutron-multiplying medium is reduced.

The studies conducted show that by changing the channel lattice parameters—that is, by reducing the carbon moderator layer—we are unable to obtain a superthermal spectrum of moderated neutrons. However, there is still the possibility of significantly influencing the spectrum of moderated neutrons by introducing burnable neutron absorbers into the heterogeneous structure.

Calculations of neutron moderation spectra were performed for the same heterogeneous structures as in the studies described above, but with the addition to the fuel channels of two neutron absorbers in the form of 1-mm-thick cadmium (Cd) and indium (In) foils, which are cylindrical in shape and located between the uranium dicarbide fuel rod and the inner wall of the fuel channel's inner tube. Fig. 10 shows the structure, composition, and geometric parameters of the fuel channel with burnable neutron absorbers, which are specified for the calculations of the spectra presented below in Fig. 11.

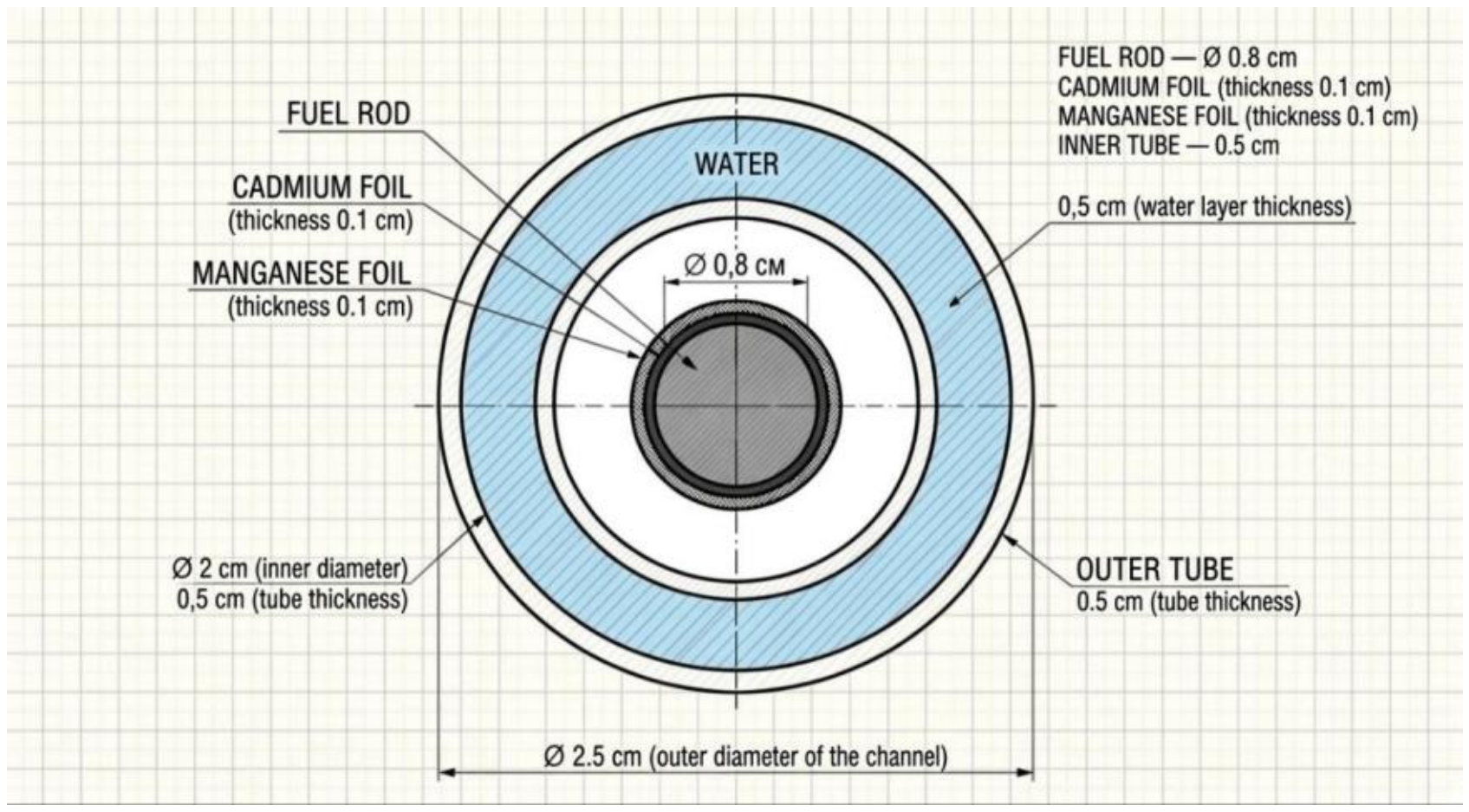

*Fig. 10. Structure, composition, and geometric parameters of the fuel channel specified for the calculations of the spectra presented below in Fig. 11.*

Fig. 11 shows the calculated moderation spectra obtained for a heterogeneous uranium-carbon channel simple lattice (Fig. 5 and Fig. 6) with a lattice parameter of 30 cm, and the fuel channel structure and geometry shown in Fig. 10, for uranium dicarbide with 5% enrichment in uranium-235.

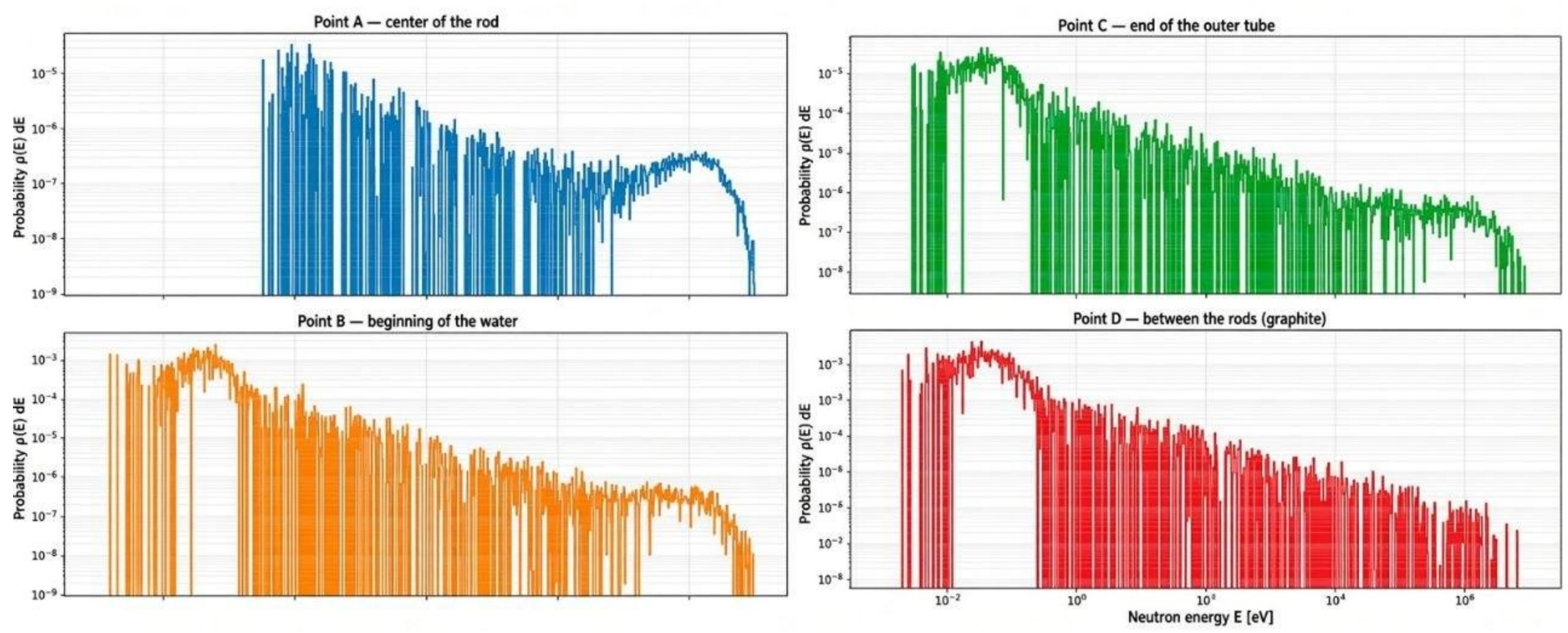

*Fig. 11. Calculated moderation spectra obtained for a heterogeneous uranium-carbon channel simple lattice (Fig. 5 and Fig. 6) with a lattice parameter of 30 cm, and the structure and geometry of the fuel channel shown in Fig. 10, for uranium dicarbide with a 5% enrichment in uranium-235.*

The top spectrum of moderated neutrons shown in Fig. 11 indicates that the maximum of the neutron spectrum in the fuel rod lies in the superthermal region of neutron energies (1–20 eV). Furthermore, a comparative analysis of the neutron spectra in Fig. 11 reveals the effect of introduced burnable neutron absorbers on the neutron spectrum.

Thus, the conducted studies make it possible to determine the structure and composition of a heterogeneous uranium-carbon fission medium, which is promising for mathematical modeling and determining the traveling-wave regime of neutron-nuclear fissions. Consequently, if the existence of such a mode is confirmed by the results of mathematical modeling, it will be possible to determine the structure and composition of the core of a nuclear reactor operating in a traveling-wave mode of neutron-nuclear fissions on superthermal neutrons (1–7 eV) in uranium reactor fuel [3].

## IV. Neutron Modulation Spectra in a Heterogeneous Thorium-Carbon Fission Medium

As noted above in Section I, the traveling-wave mode of neutron-nuclear fission in thorium fission media can be realized using thermal neutrons [6–10]; therefore, the thorium fission medium must contain a neutron moderator, i.e., it must be heterogeneous. Therefore, it is necessary to determine the possible composition and structure of such a heterogeneous thorium fission medium. To this end, this study used the OpenMC software code, which implements the Monte Carlo method, to investigate the energy spectra of moderated neutrons in heterogeneous thorium-carbon fission structures.

The simulated heterogeneous thorium-carbon fission structure was analogous to those used in the study of moderated neutron spectra in heterogeneous uranium-carbon media in Section III, and is shown in Fig. 5 and Fig. 6. It consists of a simple lattice of fuel channels 4 cm in diameter with a lattice spacing of 30 cm, arranged within a pyrocarbon block, which serves as the primary moderator for the neutrons emitted by the fuel channels. The fuel channels consist of two tubes made of structural reactor metals that weakly absorb neutrons (e.g., zirconium or aluminum alloys) and are arranged coaxially; their effect on the neutron spectrum was not taken into account. The inner tube, with a diameter of 2 cm, is filled with fissile thorium material (thorium-232 enriched to 10% in uranium-233). The volume between the inner and outer tubes of the fuel channel was assumed to be filled with water during the calculation, which corresponds to the heat removal system of an RBMK reactor.

Fig. 12 shows the calculated neutron moderation spectra for several points in a heterogeneous thorium-carbon fission medium, which has the same geometric structure as that shown in Figs. 5 and 6. Point 1 (Fig. 6) is a point located in the midplane of the heterogeneous structure and perpendicular to the fuel channels at the center of the central fuel rod; Point 2 (Fig. 6) is a point located at the interface between the fuel channel and the carbon moderator of the central rod; point 3 (Fig. 6) is a point located in the same plane as the two previous points and in the carbon medium between the two rods.

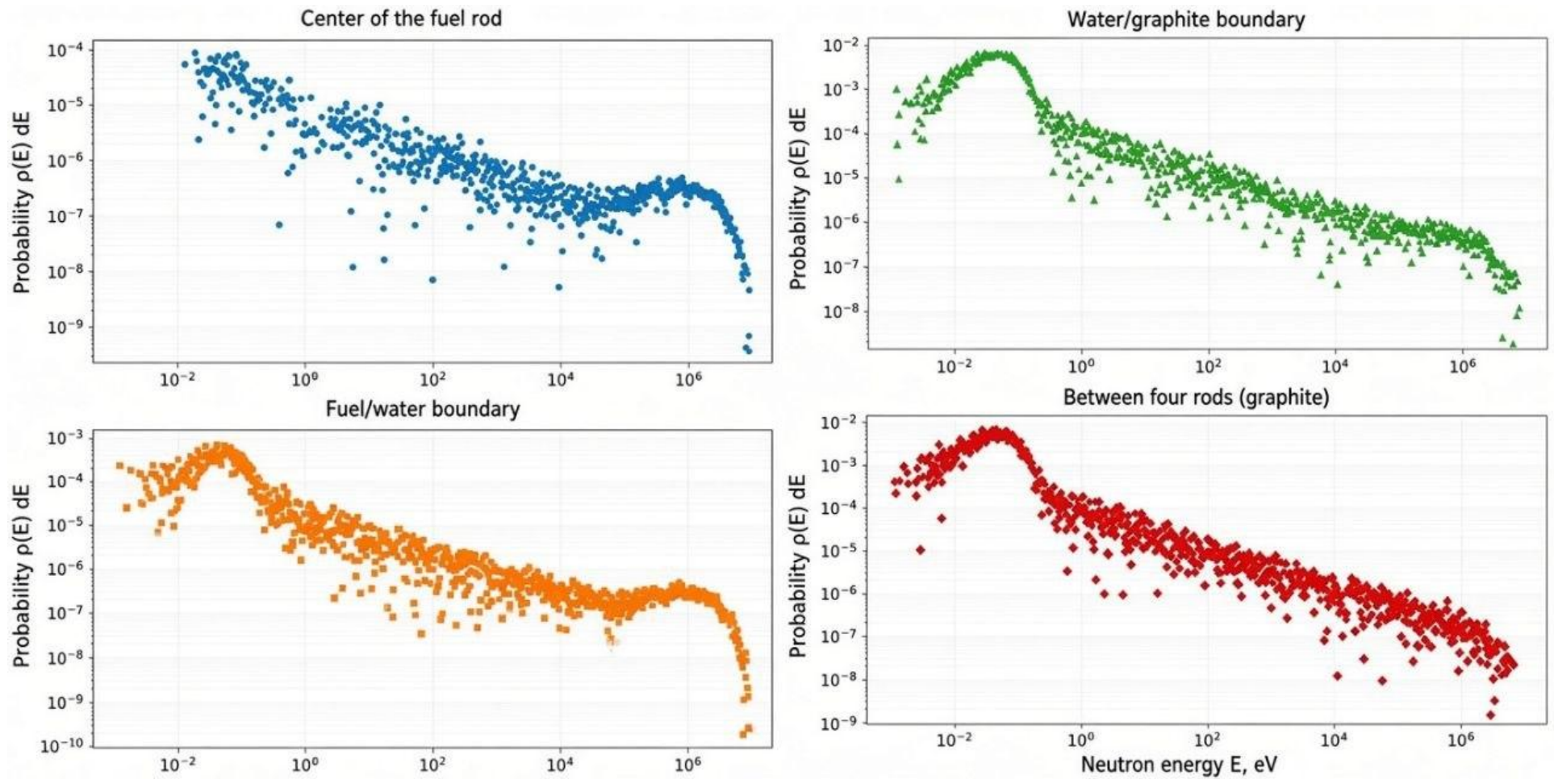

*Fig. 12. Calculated neutron moderation spectra for several points in a heterogeneous thorium-carbon fission medium having the same geometric structure as shown in Fig. 5 and Fig. 6. Point 1 (Fig. 6) is a point located in*

*the midplane of the heterogeneous structure and perpendicular to the fuel channels at the center of the central fuel rod; Point 2 (Fig. 6) is a point located at the interface between the fuel channel and the carbon moderator of the central rod; point 3 (Fig. 6) is a point located in the same plane as the two previous points and in the carbon medium between the two rods.*

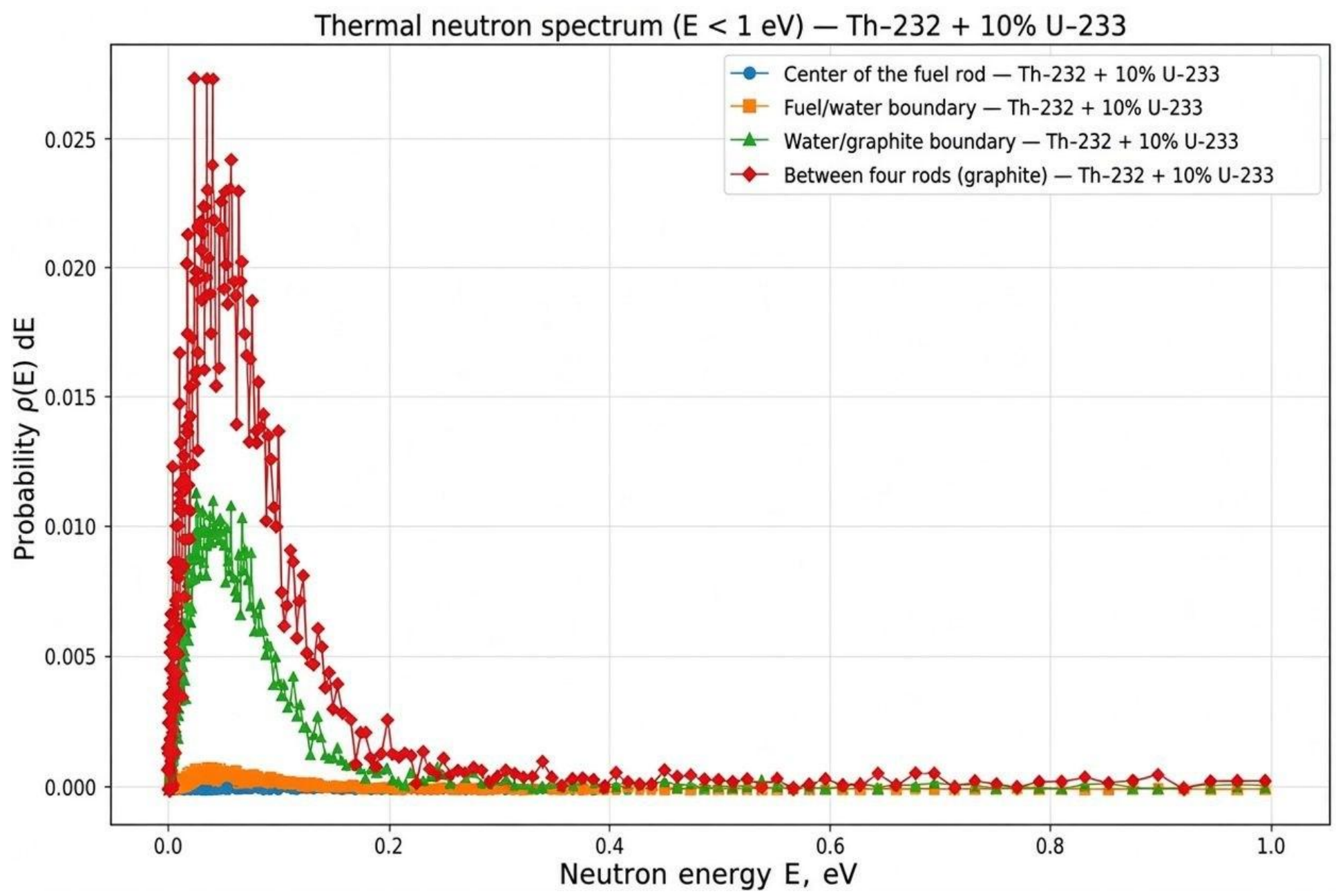

*Fig. 13. The same calculated neutron moderation spectra for several points in the heterogeneous thorium-carbon fission medium shown in Fig. 12, but for the neutron energy range of 0–10 eV.*

The results presented in Fig. 12 and Fig. 13 indicate that, in the given heterogeneous thorium-carbon fission medium, the neutron moderation spectra are thermal. Thus, a heterogeneous thorium-carbon fission medium has been identified in which the traveling-wave regime of neutron-nuclear fissions can be realized.

## V. CONCLUSIONS

In this work, using the Geant4 [11] and OpenMC [12] software codes, which implement the Monte Carlo method, a study was conducted of the energy spectra of slowed-down neutrons in various homogeneous and heterogeneous uranium and thorium fission structures with different compositions and lattice parameters.

For example, calculated neutron spectra were obtained for homogeneous uranium fission media (uranium dicarbide and uranium dioxide), and it was shown that in homogeneous uranium dicarbide, the spectral maximum lies in the 20–50 keV range. This is important for the development of a prototype fast single-channel reactor operating in a traveling-wave fission mode with a soft fast neutron spectrum.

The composition and lattice parameters of a heterogeneous channel-type thorium-carbon fission medium have been determined; in this medium, a thermal spectrum of slowed-down neutrons is formed, making it possible to achieve a traveling-wave mode of neutron-induced nuclear fission on thermal neutrons.

In addition, the spectra of slowing neutrons in heterogeneous uranium-carbon (carbon as the neutron moderator) channel-type fission media were obtained for various channel lattice parameters and compositions of the fission media. During the calculations of neutron moderation spectra, a heterogeneous uranium fission structure—analogous to the structure of the heterogeneous core of an RBMK channel-type nuclear reactor—was used to verify the correctness of the simulation program's computational scheme. Analysis of the obtained spectra confirms the correctness of the simulation program's computational scheme. The conducted research and the obtained results make it possible to determine the structure and composition of a heterogeneous uranium-carbon fissile medium, which is promising for mathematical modeling and determining the traveling-wave regime of neutron-nuclear fissions on superthermal neutrons.